# Highly efficient UV detection in a metal-semiconductor-metal detector with epigraphene

Hans He,[1,2] Naveen Shetty[1], Sergey Kubatkin[1], Pascal Stadler[1], Tomas Löfwander[1], Mikael Fogelström[1], J.C. Miranda-Valenzuela[3], Rositsa Yakimova[4], Thilo Bauch[1], and Samuel Lara-Avila[1,5,*]

[1] *Department of Microtechnology and Nanoscience, Chalmers University of Technology, 412 96 Gothenburg, Sweden*

[2] *RISE Research Institutes of Sweden, Box 857, S-50115 Borås, Sweden*

[3] *Tecnologico de Monterrey, Campus Santa Fe, 01389 Ciudad de México, Mexico*

[4] *Department of Physics, Chemistry and Biology, Linköping University, 581 83 Linköping, Sweden*

[5] *National Physical Laboratory, Hampton Road, Teddington TW11 0LW, United Kingdom*

**Abstract (250 words)**

We show that epitaxial graphene on silicon carbide (epigraphene) grown at high temperatures (T > 1850 °C) readily acts as material for implementing solar-blind ultraviolet (UV) detectors with outstanding performance. We present centimeter-sized epigraphene metal-semiconductor-metal (MSM) detectors with peak external quantum efficiency of $\eta \sim 85\%$ for wavelengths $\lambda = 250\text{-}280$ nm, corresponding to nearly 100% internal quantum efficiency when accounting for reflection losses. Zero bias operation is possible in asymmetric devices, with the responsivity to UV remaining as high as $R = 134$ mA/W, making this a self-powered detector. The low dark currents $I_o \sim 50$ fA translate into an estimated record high specific detectivity $D = 3.5 \times 10^{15}$ Jones. The performance that we demonstrate, together with material reproducibility, renders epigraphene technologically attractive to implement high-performance planar MSM devices with a low processing effort, including multi-pixel UV sensor arrays, suitable for a number of practical applications.

Detectors of ultraviolet radiation (UV, $\lambda = 10 - 400$ nm) have proven to be useful in a number of applications, including military, industrial, environmental, and health sectors.[1–3] The pursuit of high-performance UV detectors remains attractive for applications such as non-line-of-sight (NLoS) communications[4–6] and UV astronomy.[7–9] Quantum noise limited UV photon counting receivers would enable high data transfer rates on short to medium NLoS ranges, exploiting the nearly zero UV background conditions on ground level.[10] Astronomy in the UV range, on the other hand, can play a crucial role in understanding the chemical evolution of the universe, and provide evidence for dark energy in cosmology studies.[11,12] To achieve the ultimate sensitivity in the UV range, these and other applications require solar (or visible) blind detectors, which should be insensitive to light with wavelength $\lambda > 400$ nm. A preferred route to achieve solar blindness in photovoltaic solid-state detectors is by using wide bandgap semiconductors ($E_G \gtrsim 3$ eV) as the core of UV detection. A large $E_G$ imposes constraints on photon energies $E_{ph}$ for efficient absorption and electron-hole pair generation, removing the need to use optical filters in practical detectors.

Silicon carbide (SiC) is one of the most studied wide-gap materials ($E_G = 3.26$ eV for the 4H polytype at 300 K), and since the first observation of UV detection in SiC p-n junctions,[13] it has become a preferred choice for UV photodetection, providing practical advantages such as operation in harsh environments and high temperatures.[13–16] Conventional SiC UV detectors are based on UV photons being absorbed in the bulk of SiC, producing electron-hole pairs than can be separated by an electric field, and thereby generating an external current which is proportional to the number of absorbed photons. The technology of SiC-based UV detectors has been optimized over time employing different architectures[17–19] with the goal of improving the responsivity and noise performance. The responsivity $R$ quantifies the amount of photocurrent $I_{PH}$ generated under illumination at a given optical power $P_{OPT}$, with $R = I_{PH} / P_{OPT} = \eta e/(hf)$ [A/W], where $\eta$ is the external quantum efficiency, $e$ electron charge, $h$ Planck's constant, and $f$ frequency of the incoming radiation. The sensitivity of the detector is limited by noise, and the noise-equivalent-power, $NEP = S_N/R$ [W/√Hz]) provides a metric for the ultimate practically detectable signal, with $S_N$ being the current noise spectral density $S_N$ [A/√Hz]. Taking the effective area of the detector $A$ into account, the specific detectivity serves the purpose of enabling the performance comparison among different detector technologies, and it is defined as $D = \sqrt{A}/NEP$ [20], in units of Jones [cm√Hz/W].

With the advent of epitaxial growth of graphene on SiC (epigraphene), UV detection with SiC has been demonstrated using the epigraphene layer as a transparent contact.[21,22] Graphene-SiC UV detectors have been demonstrated that display responsivities of the order of 10 mA/W at $\lambda = 254$ nm, corresponding to rather modest external quantum efficiencies of $\eta \sim 4$ %.[21] Greater responsivities have been achieved at the expense of adding complexity to the devices, e.g. in vertical p-n mesa junctions formed by epilayers of SiC.[22] Yet, remaining challenges with epigraphene-based UV detectors include avoiding the formation of stacking faults in the SiC epilayers[23], avoiding the need for large bias operation[24], and minimizing dark currents.[25]

Here we present a graphene-based UV detector with peak external quantum efficiency of $\eta \sim 85\%$ for wavelengths 250 - 280 nm, corresponding to nearly 100% internal quantum efficiency when accounting for reflectance losses in the material (reported experimental reflectance for 4H-SiC is in the range 15-30%[26,27]). The device is a metal-semiconductor-metal (MSM) structure, and the high performance is based on our capability to form a reproducible junction between the high-purity semi insulating (HPSI) 4H-SiC substrate and the graphene layer grown atop. Electrically, a MSM structure is formed by two Schottky diodes connected back-to-back (Fig. 1a). Under an applied voltage, irrespective of the polarity of the bias, at least one of the diodes remains reversed-biased so that a small saturation current $I_o$ flows through the device. Upon illumination, the excess current produced by the photo-generated carriers, i.e. the photocurrent $I_{PH}$, can exceed $I_o$ by many orders of magnitude making it easily detectable. In our case, the saturation currents in dark conditions can be as low as $I_o \approx 50$ fA, and the optical power from a UV LED source can produce $I_{PH} = 1$ µA (on/off ratio > $10^7$). A practical advantage of MSM photodetectors is the simplicity in manufacture, and in our case the high performance of the devices has been observed on substrates just after growth of epigraphene, by making contacts to SiC and graphene using wedge bonding, without any need for microfabrication. However, for careful characterization, the results presented in this work were collected on devices having a well-defined geometry achieved by standard e-beam lithography. In total, two graphene substrates, grown on SiC from different commercial suppliers, were studied in detail by quantifying the responsivity, effective area, noise and speed of response.

Epigraphene was grown on the silicon-terminated face of HPSI 4H-SiC substrates (Cree Inc. and Norstel AB) by encasing them in a graphite crucible, in an inert argon atmosphere of 850 mbar, and heating using RF power. We have optimized the growth temperatures T > 1850 °C using a design of experiments (DOE) methodology[28]. Transmission mode microscopy was used to quantify the content of bilayer graphene in the range of 10 - 20% for the substrates used in our experiments.[29] Device fabrication used standard electron beam lithography techniques. Epigraphene was patterned using oxygen plasma etching and metal contacts were deposited using physical vapor deposition of 5 nm Ti and 80 nm Au.[30] Optical characterization of the substrates and spectral responsivity measurements were taken using an Agilent Cary 60 spectrophotometer in the range 180 nm - 1100 nm. Electrical characterization was performed under ambient illumination conditions at room temperature using a Variable Gain Sub Femto Ampere Current Amplifier DDPCA-300, and a Variable Gain High-Speed Current Amplifier DHPCA-100 from Femto Messtechnik GmbH. The samples were illuminated in two different ways: 1) by a deep UV (DUV) flood exposure unit from Bachur & Associates, equipped with a DUV 220 nm mirror, primarily reflecting radiation with $\lambda$ = 220 - 260 nm, providing an optical power density up to p=10 mW/cm$^2$, 2) a continuous LED light source ($\lambda$ = 250 nm) LLS-250 from Ocean Optics fiber-coupled to a ball lens to provide a light spot of diameter about $D \sim 1.5$ mm, providing an optical power up to $P = 6$ µW ($p = 0.3$ mW/cm$^2$). In all cases, the light intensity was calibrated using both a silicon photodiode power sensor S130VC from Thorlabs, and an optical power meter OE-200-UV from Femto Messtechnik GmbH.

Figure 1b shows the schematic of the measurement setup for Device A, which consist of a 5 x 5 mm$^2$ epigraphene square. Electrical contacts are placed around the epigraphene perimeter, with additional contacts also deposited directly onto the SiC substrate. A large area sample ensure that the UV light-spot falls entirely inside the epigraphene area during optical characterization so that all the light, except for the reflection losses, is coupled to the device. Fig. 1c shows the current-voltage (I-V) characteristic for this device both in dark and under illumination with $\lambda$ = 250 nm.

The dark current for this square geometry device is below pA in the bias range $V_B = \pm 10$ V, being as low as $I_{OFF}$ = 52 fA for $V_B$ = 10 V (epigraphene as anode). In the absence of light, the zero-bias differential resistance measured between epigraphene and the contacts on SiC is $R_0 = dV/dI \sim 20$ TΩ.

Under illumination ($P_{opt}$ = 6 µW), the current measured at $V_B$ = 10 V is $I_{ON}$ = 0.8 µA, yielding a current on-off ratio $I_{ON}/I_{OFF} \sim 1.54 \times 10^7$. The shape of the I-V curve is that of a MSM photodetector with asymmetric Schottky contacts, where the epigraphene is the anode and Ti/Au is the cathode. In the light on case, the IV curve clearly exhibit two distinct operating regions, referred as photoconductive and saturation regimes[31]. The photoconductive regime is the low forward bias with a steep sloped region of the I-V curve, while the saturation at high bias is due to all photogenerated carriers being swept out by the electric field.

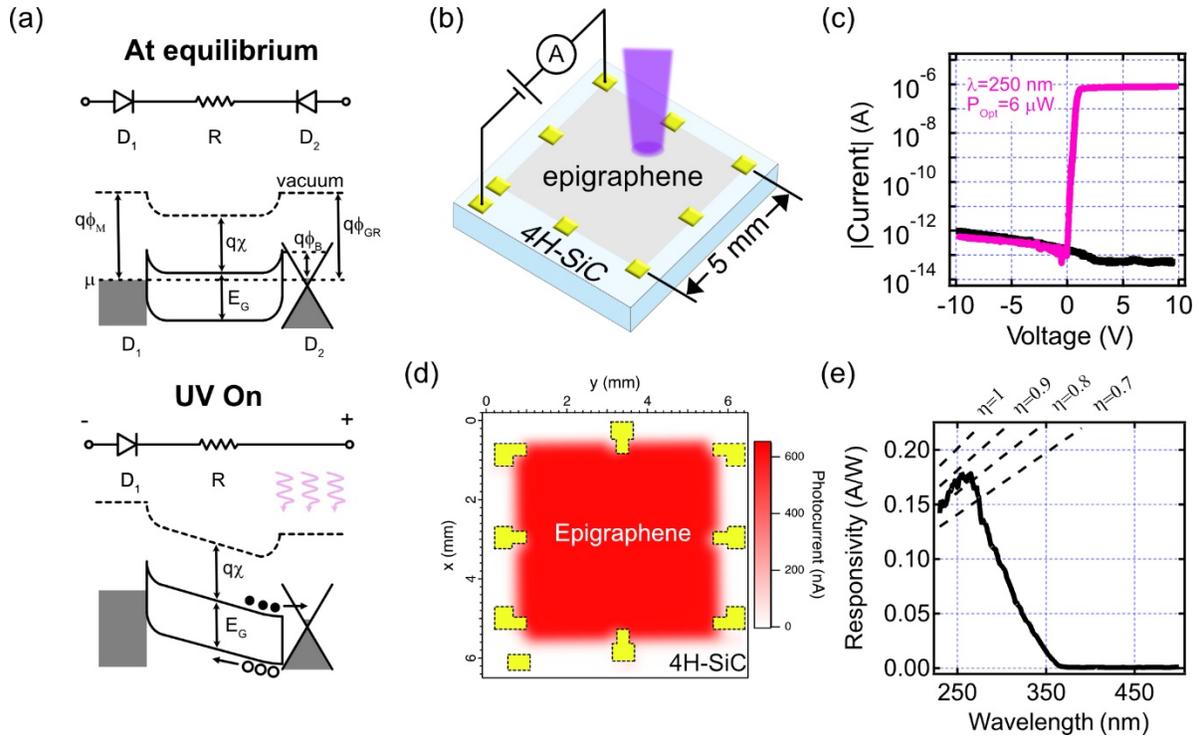

**FIG. 1.** UV detection with epigraphene (a) The metal-semiconductor metal (MSM) detector consists of two Schottky diodes (D1 and D2) connected back-to-back through the SiC channel resistance R. The energy band diagrams of semi-insulating 4H-SiC in contact with metal and graphene show the Schottky barriers formed at the left and right interface of SiC. In equilibrium, in dark condition and zero applied voltage, the interfaces form a depletion type of contact. For graphene (D2), the barrier height is $\Phi_B = \Phi_{Gr} - \chi_{4H}$- $\chi_{4H\text{-}SiC}$, where $\Phi_{Gr}$ is the work function of graphene and $\chi_{4H\text{-}SiC}$ the electron affinity of 4H-SiC. When the graphene interface is illuminated with UV-light, absorbed UV photons create electron-hole pairs via interband transitions. The carrier photogeneration is accompanied by a reduction of the Schottky barrier and under forward bias free electrons are collected on the graphene contact. (b) Schematic of device A. The device consists of an epigraphene square ($L$ = 5 mm x $W$ = 5 mm) with eight contacts, in addition to a Ti/Au contact deposited directly on 4H-SiC. (c) I-V characteristic under illumination (pink) and under dark conditions (black). (d) Scanning photocurrent at $V_B$ = 10 V shows that the effective area for photocurrent generation is right under the epigraphene sheet. (e) Spectral responsivity at $V_B$ = 10 V, when the sample is illuminated only on the epigraphene sheet (light spot 1 mm x 1.5 mm). The highest responsivity $R$ = 175 mA/W is reached at $\lambda$ = 254 nm. Dashed lines indicate different quantum efficiencies $\eta$.

A quantitative analysis of the device responsivity to UV photons requires knowledge of the active area of the device, to quantify the optical power available for photocarrier generation. The effective area of the device was investigated by scanning the light-spot across the entire chip, while keeping a constant bias $V_B$ = 10 V. Fig. 1d shows that the effective photo-detecting area unambiguously corresponds to the region where the epigraphene contact covers the surface of SiC.

Having established the effective area, the spectral responsivity of the device was studied over the wavelength range $\lambda$ = 200 - 1000 nm. Fig. 1e shows the responsivity peaking at $\lambda$ = 250 – 280 nm, with a maximum measured value about $R$ = 175 mA/W, corresponding to an external quantum efficiency $\eta$ ~ 85 %. The responsivity is drastically reduced for $\lambda$ > 400 nm (*see supplementary S1*), owing to the large bandgap of 4H-SiC, and results in a visible rejection ratio of at least 1400, which we define as the ratio of the responsivity at $\lambda = 250$ nm to that at $\lambda = 400$ nm.

Of practical relevance is that the responsivity of the device remains large even at zero bias, making this a self-powered detector. Figure 2a shows the I-V response of the detector for different UV intensities. The linearity of photocurrent in the device extracted from these measurements is shown in Fig. 2b for two different biases. Typical for MSM detectors, a larger forward bias leads to a higher responsivity, due to an increase in size of the depletion width of the reverse-biased contact. The zero bias operation is possible thanks to the asymmetric contacts, i.e., contacts with dissimilar work functions, which give rise to a built-in bias in the detector in the absence of externally applied voltage. (*for device B with symmetric Schottky barriers see Supplementary S2*).

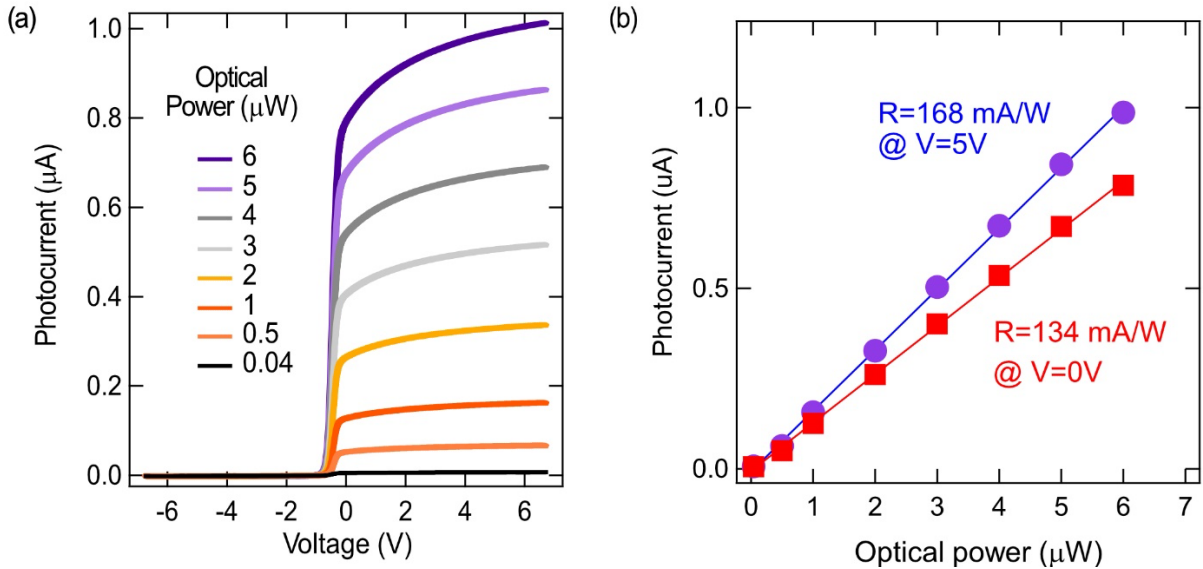

**FIG. 2.** Photo response at different UV power and bias. (a) I-V characteristics at different optical powers. (b) The extracted photocurrent at $V_B = 5$ V versus optical power shows a linear dependence, and the slope of the curve provides an alternative estimate of the responsivity $R = 168$ mA/W, which agrees well with the measurement from the spectrometer shown in the main text.

We explore the device performance in terms of speed and detectivity (Fig. 3). The rise time shown in Fig 3a is characterized by a double exponent, with $\tau_1 = 10$ μs, and $\tau_2 = 144$ μs. The fast time constant is associated with electron-hole generation process and the slow one is associated with charge redistribution processes[31]. We note that the fastest rise time was limited by the speed of the optical chopper available in our setup, so that graphene based MSM can in principle be much faster than here described (Supplementary S3).

As for detectivity, the small dark currents due to the large resistance in the dark state, $R_0 = dV/dI \sim 20$ TΩ (Fig. 1c), give expectation for large detectivity values. A common way to estimate detectivity is to assume that the noise is of purely thermal origin, which allows for an estimate of detectivity, $D \approx e\lambda\eta/hc(4k_BT/R_0A)^{-1/2}$ ([17,20]). If we apply this formula with $\lambda = 250$ nm, T = 300 K, $\eta = 1$, $R_0 = 20$ TΩ, and $A = 5$ mm x 5 mm, the detectivity is $D = 3.5 \times 10^{15}$ Jones, which is among the highest theoretical estimates ever reported in literature for UV detectors[17]. Yet, we can compare this estimate with experimental data. For this, Fig. 3b shows the noise characterization of the device. The current spectral noise density is flat, $S_N = 0.7$ pA/√Hz above $f \sim 10$ Hz. At optimal responsivity of $R = 168$ mA/W, this leads to a $NEP = S_N/R$

= 4.2 pW/√Hz, which in turn results in a detectivity of about $D = \sqrt{A}/NEP = 1.2 \times 10^{11}$ Jones for Sample A with area $A = 5 \times 5$ mm$^2$. This is the experimentally determined detectivity, limited by limited by the amplifier noise, whereas theoretically one should achieve 4 orders of magnitude better detectivity.

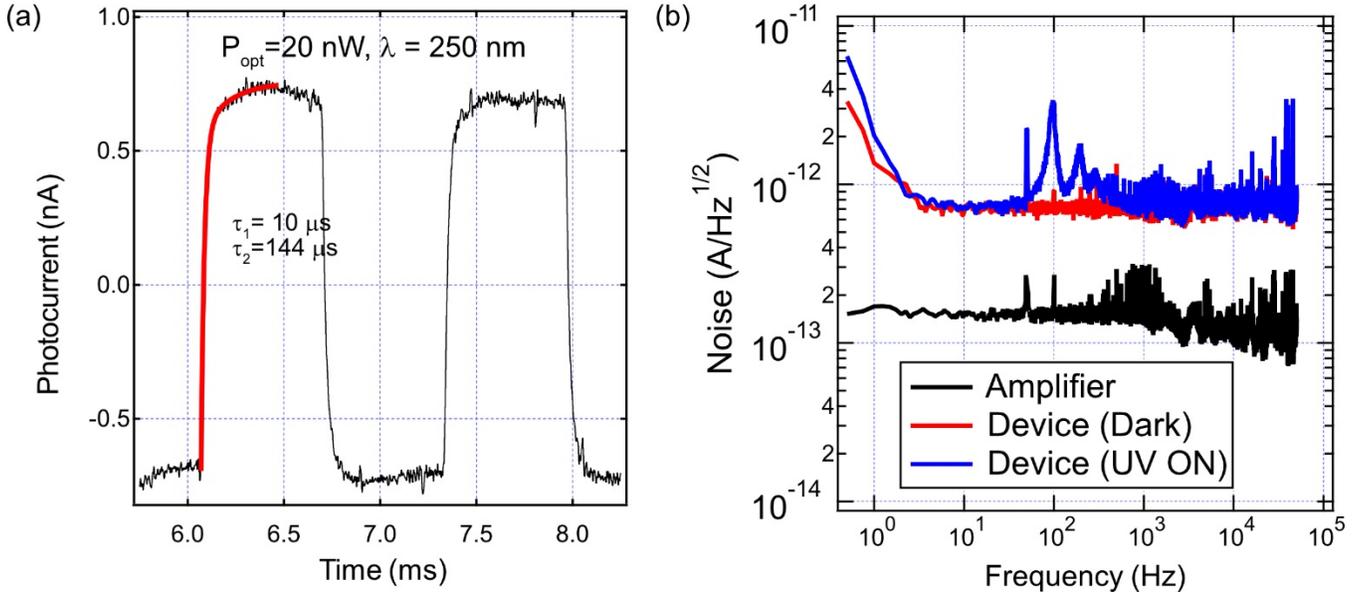

**FIG. 3.** UV Detector performance of epigraphene. (a) Rise time and fall time for square device (Sample A, Fig. 1). Rise times are fitted with a double exponent (red), yielding two time scales $\tau_1 = 10$ μs and $\tau_2 = 144$ μs. (b) Noise current spectral density versus frequency. The amplifier noise (black) was measured in open circuit conditions. The noise floor of our setup is limited by the amplifier noise of 0.2 pA/√Hz.

Finally, we mention that the unprecedently high responsivity to UV light demonstrated here for the epigraphene system is observed only in graphene grown at high temperatures, T > 1850 °C (Supplementary S4). It is known that the recombination lifetime of photo-generated carriers in 4H-SiC can be increased by annealing at high temperatures ( > 1700 °C) [32–36]. Thus, we suggest that the high temperature process needed to produce epigraphene with high crystalline quality is the cause for a reduced concentration of defects at the graphene-SiC substrate interface that could act as charge traps and/or recombination centers. A typical challenge of MSM detectors has been to reliably control the metal-semiconductor interface. For epigraphene, the graphene layer acting as the metal is produced at high temperature in quasi-thermodynamic conditions, making it highly reproducible. Epigraphene is particularly suitable to implement the fully planar device architecture of the MSM structure with a low processing effort and makes it technologically attractive for the development of high-performance multi-pixel UV sensor arrays.

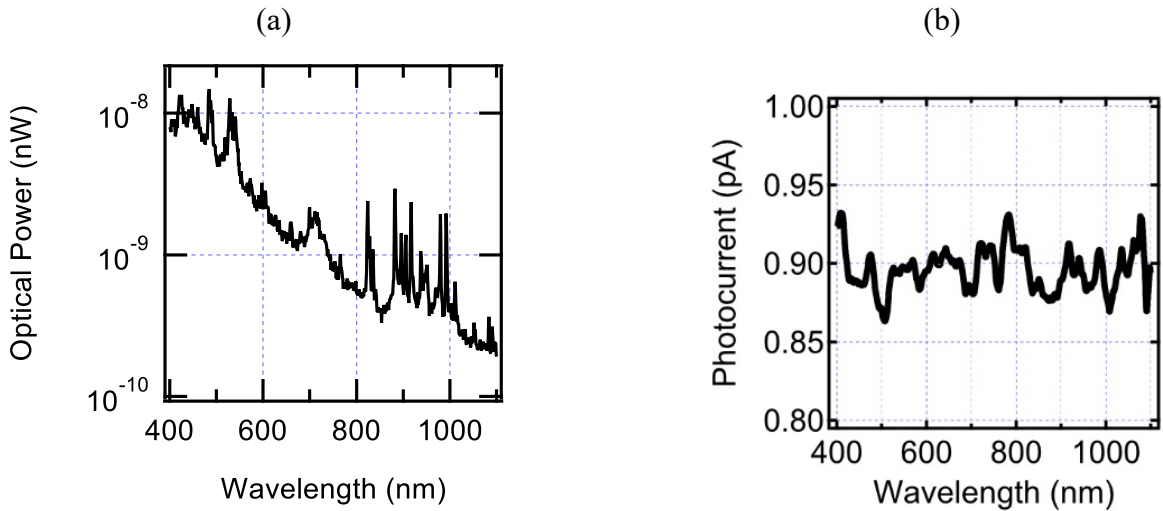

**Figure S1. Responsivity in the range $\lambda$ = 400 - 1100 nm.** (a) Optical power of the Agilent Cary 60 photo spectrometer in the range $\lambda$ = 400 – 1100 nm, measured with a silicon photodiode power sensor S130VC from Thorlabs. (b) The photocurrent measured for the epigraphene detector in the range $\lambda$ = 400 – 1100 nm is constant and just below pA, implying that the responsivity is drastically reduced for $\lambda$ > 400 nm, owing to the large bandgap of 4H-SiC. Together with the optical power of a source, an estimation of the responsivity at $\lambda$ = 400 nm is $R_{400nm} = I_{PH} / P_{OPT} =$ 0.9 pA/ 10nW = 90μA/W, and we conclude that the epigraphene-based photodetector has a visible rejection ration of at least $R_{250nm}/R_{400nm} =$ (175 mA/W)/(90 μA/W)=1944, which we define as the ratio of the responsivity at 250 nm to the responsivity at 400 nm.

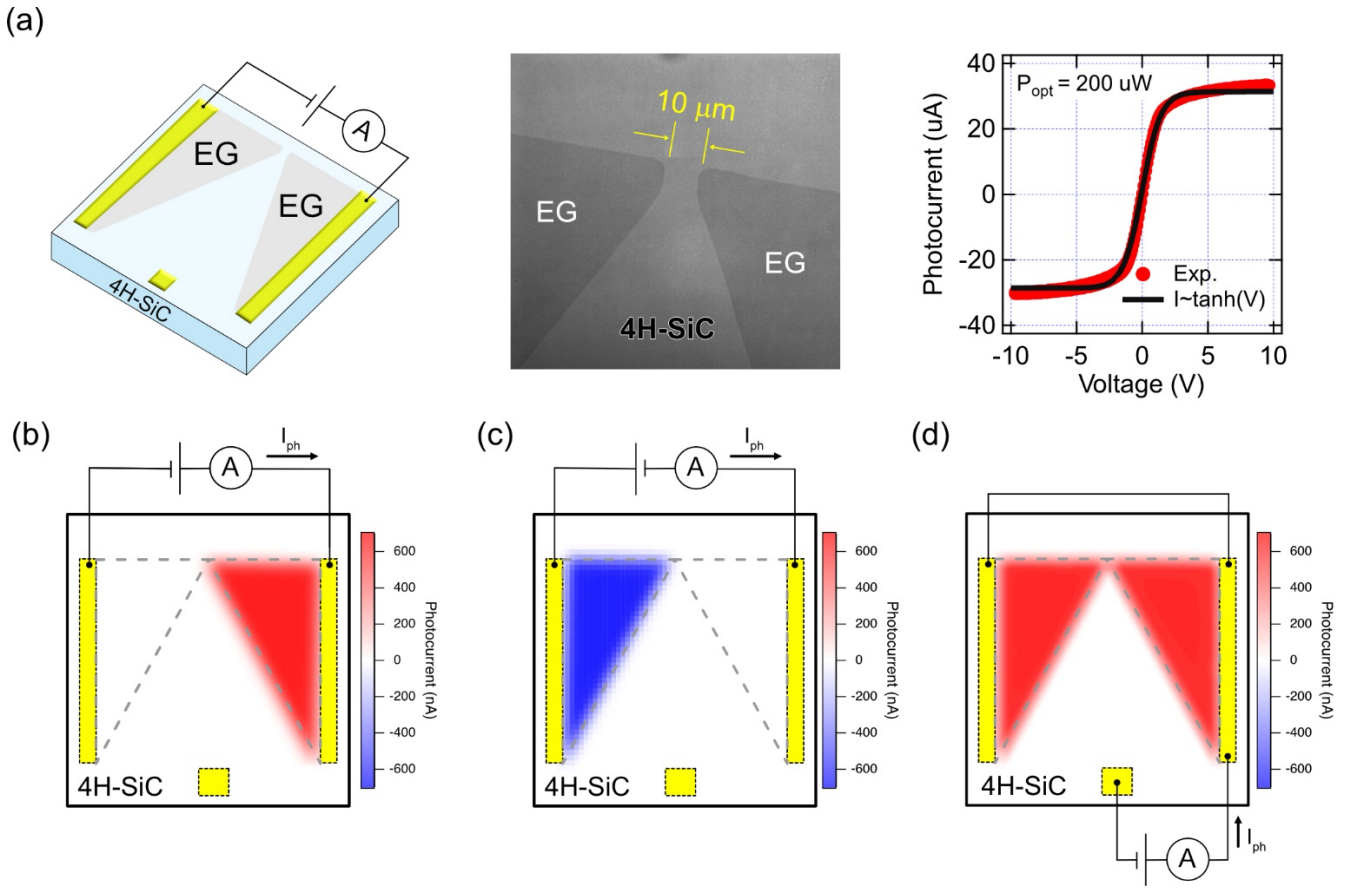

**FIG. S2. Sample B.** (a) Schematic of sample B, which consists of two triangular-shaped epigraphene contacts, separated by a gap of 10 μm. There is an additional Ti/Au contact deposited directly on 4H-SiC. The I-V characteristic shows that for symmetric contacts and even illumination of both contacts, the magnitude of the photocurrent is the same for positive and negative bias, $|I_{Ph}(V)| = |I_{Ph}(-V)|$. (b) Scanning photocurrent measurements ($P_{OPT}$ = 3 μW) with positive polarity applied to the right contact. Only the graphene contact connected to the positive terminal of the voltage source produces a photocurrent when illuminated with DUV. (c) If the polarity is inverted, photocurrent is produced when the left contact is illuminated with UV. (d) With the two graphene electrodes connected and held at the same positive bias, both graphene contacts produce photocurrent when illuminated with DUV (the metal contact on SiC is the cathode).

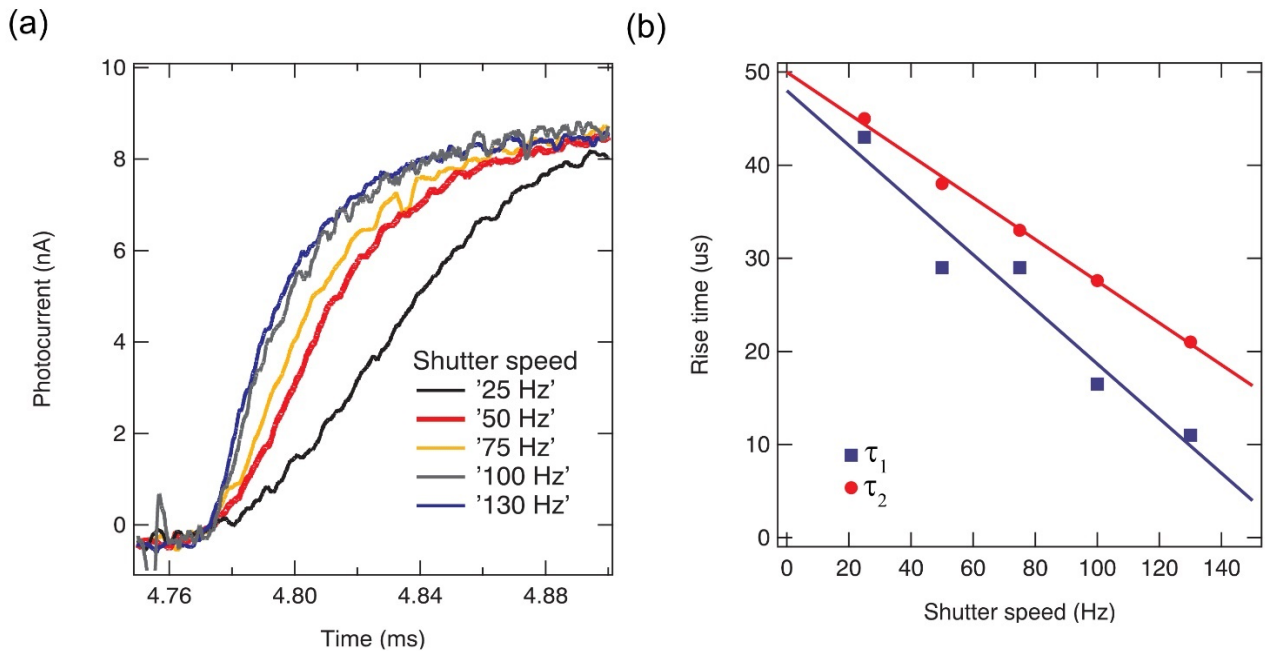

**Figure S3. Speed of the epigraphene MSM UV detector.** (a) Speed of response (rise time) of photocurrent measured at different shutter speeds. (b) The rise times extracted by a two-exponent fit. The device speed seemingly increases linearly with shutter speed (shorter rise times), up to the maximum tested frequency of 130 Hz: measurements of rise times are limited by the set-up.

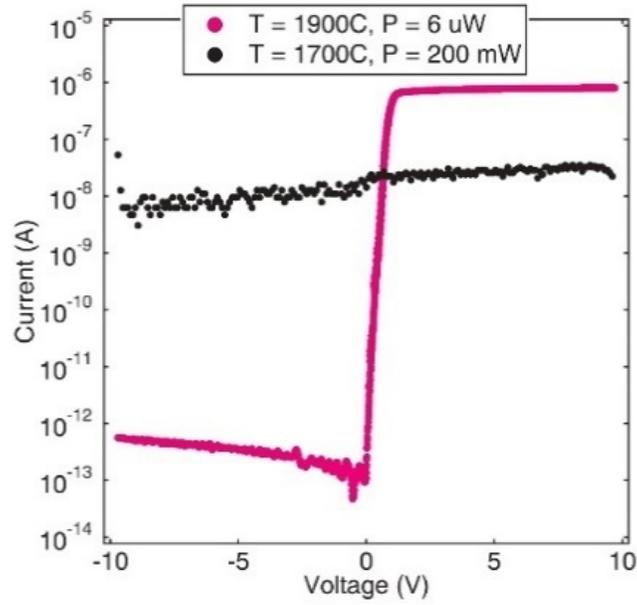

**Figure S4. Comparison of UV detection in epigraphene grown at different temperatures.** Representative example of samples grown at T = 1700 °C, in which the photocurrent reaches only $I_{PH} \sim$ 20 nA in the entire bias range (black dots), despite being flood illuminated with a UV lamp. The photocurrent measurement for sample A, grown at T = 1900 C and illuminated with UV LED source is added for comparison.

## ACKNOWLEDGMENTS

This work was jointly supported by the Swedish Foundation for Strategic Research (SSF) (nos. GMT14-0077, RMA15-0024), Chalmers Excellence Initiative Nano, 2D TECH VINNOVA competence Center (Ref. 2019-00068). This work was performed in part at Myfab Chalmers. The authors declare that the main data supporting the findings of this study are available within the article and supplementary information. Additional data are available from the corresponding author upon request.